\documentclass{article}
\usepackage[T1]{fontenc} 
\usepackage[utf8]{inputenc} 
\usepackage{ismir,amsmath,cite,url}
\usepackage{graphicx}
\usepackage{color}
\usepackage{amssymb}
\usepackage{mathrsfs}
\usepackage{mathabx}
\usepackage[bookmarks=false,hidelinks]{hyperref}

\usepackage{multirow}

\newcommand*\BitAnd{\mathrel{\&}}
\newcommand*\BitOr{\mathrel{|}}
\newcommand*\ShiftLeft{\ll}
\newcommand*\ShiftRight{\gg}

\newcommand*\BitXor{\mathrel{\vee}}

\title{Quantifying Musical Style: Ranking Symbolic Music based on Similarity to a Style}

\twoauthors
  {Jeff Ens} {Simon Fraser University \\ {\tt jeffe@sfu.ca}}
  {Philippe Pasquier} {Simon Fraser University \\ {\tt pasquier@sfu.ca}}

\begin{document}

\maketitle
%


\begin{abstract}
Modelling human perception of musical similarity is critical for the evaluation of generative music systems, musicological research, and many Music Information Retrieval tasks. Although human similarity judgments are the gold standard, computational analysis is often preferable, since results are often easier to reproduce, and computational methods are much more scalable. Moreover, computation based approaches can be calculated quickly and on demand, which is a prerequisite for use with an online system. We propose StyleRank, a method to measure the similarity between a MIDI file and an arbitrary musical style delineated by a collection of MIDI files. MIDI files are encoded using a novel set of features and an embedding is learned using Random Forests. Experimental evidence demonstrates that StyleRank is highly correlated with human perception of stylistic similarity, and that it is precise enough to rank generated samples based on their similarity to the style of a corpus. In addition, similarity can be measured with respect to a single feature, allowing specific discrepancies between generated samples and a particular musical style to be identified.
\end{abstract}
\section{Introduction}\label{sec:introduction}

Measuring musical similarity is a fundamental challenge, related to many tasks in Music Information Retrieval (MIR). In this paper, we focus on measuring the similarity between a MIDI file and an arbitrary musical style. In a musical context, the term style can refer to historical periods, composers, performers, sonic texture, emotion, and genre \cite{Dannenberg2010}. Here, we use the term style to denote the musical characteristics exhibited by a corpus $\mathcal{C} = \{C_1, ..., C_n\}$, as expressed by a feature set $\mathcal{F}$. Depending on the contents of $\mathcal{C}$, style may correspond to something as specific as a subset of a composer's work, as general as the entirety of Western Classical Music, or as personal as the musical preferences of an individual.


We propose StyleRank\footnote{The code is available at \href{https://github.com/jeffreyjohnens/style\_rank}{https://github.com/jeffreyjohnens/style\_rank}}, a method for ranking MIDI files based on their similarity to a style delineated by $\mathcal{C}$. It can be used as a tool for musicological research, to evaluate Style Imitation (SI) systems, and to filter the output of an SI system. An SI system aims to generate music that exhibits the stylistic characteristics of $\mathcal{C}$ \cite{Pasquier2017}. The primary contributions are as follows: a collection of novel features for symbolic music representation; an efficient MIDI feature extraction tool written in C++ with bindings in Python; a measure of similarity with respect to an arbitrary style delineated by $\mathcal{C}$; and two experiments demonstrating that this measure is robust, and highly correlated with human perception of stylistic similarity.

\section{Motivations}




There are several motivating factors for this research. In general, modelling human perception of musical similarity is of particular interest within the areas of Musicology, Music Cognition, and Music Theory \cite{volk2016music}. Moreover, robust measures of musical similarity are critical for many MIR tasks, including database querying, music recommendation, and genre recognition. Although human perception is the gold standard for measuring musical similarity, natural human limitations place restrictions on the quantity and speed at which judgments can be collected, directly motivating automated measures of musical similarity.

More specifically, there are inherent challenges in designing a robust and reproducible listening experiment to evaluate SI systems. There are many variables which directly effect the quality of an experimental result, such as the number of participants, the listening environment, the sound equipment, and the number of samples selected for comparison. Even controlling for those variables, there is significant variability in how music is perceived, based on one's level of training \cite{besson2007} and musical background \cite{hannon2005,Prince2009,eerola2006}, which can result in a limited inter-rater agreement \cite{schedl2013}. This is a particular issue, as it may hamper reproducibility and comparison with previously published results.


In most cases, sampling from an SI system is a stochastic process, and as a result, generated samples vary in quality. Developing a filtering process for generated material is a high priority concern, as low quality samples are undesirable when using a generative model in a production setting. Although measuring the log-likelihood of a sample can be useful as a proxy for quality, there are cases where log-likelihood significantly diverges from human perception. Theis et al. provide examples of generated images with high log-likelihood and extremely low quality \cite{Theis2015d}. To the best of our knowledge, there are no pre-existing methods for ranking generated samples with respect to an arbitrary style.

\section{Related Work}





A wide variety of similarity measures have been developed to measure melodic \cite{velardo2016symbolic}, harmonic \cite{pickens2002harmonic,paiement2005probabilistic,deHaas2013} and rhythmic similarity \cite{toussaint2004comparison}. Many of these algorithms measure similarity by comparing two symbolic sequences \cite{volk2012towards}. Stylistic similarity, however, is rarely exhibited through sequence similarity, but rather through the repeated use of particular musical devices (i.e. melodic phrases, voice leading, and chord voicing) interspersed throughout the material \cite{volk2012towards}. In order to address this concern, approaches based on compression or pattern extraction have been proposed to measure similarity \cite{liu2005efficient,ahonen2011compression,boot2016evaluating}. Since we aim to measure similarity with respect to $\mathcal{C}$, a more suitable approach will leverage information about the discriminative aspects of the entire corpus $\mathcal{C}$, rather than only taking two MIDI files into consideration.



In the context of SI system evaluation, the Turing Test \cite{Turing1995} and the Consensual Assessment Technique \cite{amabile1982} have been used to measure the stylistic similarity between generated artifacts $\mathcal{G} = \{\mathcal{G}_1,...,\mathcal{G}_m\}$ and a particular style $\mathcal{C}$ \cite{liang2017,pearce2007}. Objective measures have also been used to evaluate SI systems. Dong et al. measure the ratio of empty bars, pitch class diversity, note duration, rhythmic consistency, and tonal distance \cite{dong2018musegan}. Trieu and Keller propose a variety of metrics ranging from rhythmic variety to harmonic consistency \cite{Trieu2018}. Since these metrics produce a single scalar value, it is easy to compare $\mathcal{C}$ and $\mathcal{G}$. However, these high-level metrics are likely only capable of measuring stylistic similarity in a very general sense. Sturm and Ben-Tal. plot distributions of meter, mode, number of tokens, pitch and pitch class for $\mathcal{C}$ and $\mathcal{G}$, but do not provide an automated method for analyzing discrepancies \cite{Sturm2017}.

More comprehensive methodologies have been proposed, which involve computing all pairwise inter-set distances between samples in $\mathcal{C}$ and $\mathcal{G}$ $(D_{\mathcal{C} \mathcal{G}} = [\texttt{dist}(c,g) : (c \in \mathcal{C}) \land (g \in \mathcal{G})])$, as well as all pairwise intra-set distances for samples within a set $(D_{\mathcal{G} \mathcal{G}} = [\texttt{dist}(g_i,g_j) : (g_i \in \mathcal{G}) \land (g_j \in \mathcal{G}) \land (g_i \neq g_j)])$. \footnote{Note that we adapt the set-builder notation to construct a list (e.g., $[i/2 : 0 \leq i < 4] = [0,0,1,1]$), which unlike a set, may contain duplicate values.} CAEMSI \cite{Ens2018}, a domain independent framework for the analysis of SI systems, provides a statistical method to test the null hypothesis $H_0 : (D_{\mathcal{G} \mathcal{G}} \neq D_{\mathcal{C} \mathcal{G}}) \lor  (D_{\mathcal{C} \mathcal{C}} \neq D_{\mathcal{C} \mathcal{G}}) \lor  (D_{\mathcal{C} \mathcal{C}} \neq D_{\mathcal{G} \mathcal{G}})$ against the alternative hypothesis $H_1 : D_{\mathcal{G} \mathcal{G}} = D_{\mathcal{C} \mathcal{C}} = D_{\mathcal{C} \mathcal{G}}$. Yang and Lerch extract multi-dimensional features from each MIDI file \cite{Yang2018}. For each feature, $D_{\mathcal{C} \mathcal{G}}$ and  $D_{\mathcal{C} \mathcal{C}}$ are constructed using Euclidean distance and smoothed using kernel density estimation \cite{rosenblatt1956,parzen1962}. The distance between $D_{\mathcal{C} \mathcal{C}}$ and $D_{\mathcal{C} \mathcal{G}}$ is measured using (1) the area of overlap and (2) the Kullback–Leibler Divergence \cite{kullback1951information}. In contrast to both of these approaches, which involve evaluating the similarity between $\mathcal{G}$ and $\mathcal{C}$, StyleRank is optimized to evaluate the similarity of a single sample $g \in \mathcal{G}$ to $\mathcal{C}$.

\section{Features}



Although the features extracted by jSymbolic2 \cite{mckay2018} are quite comprehensive, many features are high-level, and thus, ill-suited for the fine-grained distinctions that are necessary to rank stylistically similar MIDI files. For example, the \texttt{Chord Type Histogram} feature contains only 11 categories. In order to capture the complexity of the musical material being analyzed, we extract a variety of high-dimensional categorical distributions from a single MIDI file. A categorical distribution is a discrete probability distribution describing a random variable that has $k$ possible distinct states. In what follows we adopt the following notation. Given a set $x$, $||x||$ denotes the number of elements in the set $x$, $\min(x)$ and $\max(x)$ denote the minimum and maximum element in $x$ respectively, and $x_i$ denotes the $i^{th}$ element in $x$. $x \setminus y$ is the set difference between $x$ and $y$, and $x \bigtimes y$ is the Cartesian product of $x$ and $y$. $\ShiftLeft{}$ indicates a left bitwise shift and $\ShiftRight{}$ indicates a right bitwise shift. $\BitAnd{}$, $\BitXor{}$, and $\BitOr{}$ refer to the bitwise $\texttt{AND}$, $\texttt{XOR}$, and $\texttt{OR}$ operations, respectively.

\subsection{Pitch Class Set Representations}



In order to reduce the number of chords, we discard octave information and represent chords as pitch class sets, using a $12$-bit integer to denote the presence or absence of a particular pitch class $(\text{C}=0,\text{C\#}=1,...,\text{B}=11)$. For example, the C-major chord $\{60, 64, 67\}$ corresponds to the pitch class set $x = \{0, 4, 7\}$, which corresponds to the integer $\sum_{i=1}^{||x||} (1 \ShiftLeft{} x_i) = 2^0 + 2^4 + 2^7 = 145$. Since there are $12$ pitch classes, there are $2^{12} = 4096$ pitch class sets, which greatly reduces the possible number of chords. However, it is possible to further reduce this space if we create an equivalence class for all transpositionally equivalent pitch class sets. For example, the pitch class sets $\{0, 4, 7\}$ and $\{2, 5, 10\}$ are transpositionally equivalent, as both are major chords, the only difference being their root. This results in $352$ distinct pitch class sets (PCD). Using Eq. \eqref{rollc} a PCD can be calculated, where $x$ is an 12-bit integer. Notably, pitch class sets are considered equivalent under the reversal operation when calculating the Forte number of a pitch class set \cite{forte}. Consequently, the pitch class sets $\{0,4,7\}$ and $\{0,3,7\}$ have the same Forte number, but correspond to different PCD's.

\begin{subequations}
    \label{roll}
    \begin{align}
    \texttt{rot}(x,n,i) &=\,(x \ShiftLeft{} i) \BitOr (x {\ShiftRight{}} (n {-} i)) \BitAnd (2^n {-} 1) \label{rolla} \\
    \texttt{reduce}(x,n) &= \min(\{\texttt{rot}(x,n,i) : 0 \leq i < n\}) \label{rollb} \\
    \texttt{pcd}(x) &= \texttt{reduce}(x,12) \label{rollc}
    \end{align}
\end{subequations}

Alternatively, a pitch class set $x$ can be represented as the set of scales which are supersets of $x$. Given a scale $\mathbb{S}$, let $\mathbb{S}_i = \{(s + i) \mod 12 : s \in \mathbb{S}\}$. The scale representation can be calculated with Eq. \eqref{scale_rep}, where $\mathbb{S}^{\texttt{M}} = \{0,2,4,5,7,9,11\}$ and $\mathbb{S}^{\texttt{H}} = \{0,2,3,5,7,8,11\}$ denote the major and harmonic minor scales respectively. $\phi(\cdot)$ returns $1$ if the predicate $\cdot$ is true and $0$ otherwise.




\begin{equation}
\medmuskip=0mu
\thinmuskip=0mu
\thickmuskip=.5mu
\texttt{sc}(x) = \big(\sum_{i=1}^{12} \phi(x \subseteq   \mathbb{S}^{\texttt{M}}_i) \ShiftLeft{} i\big) + \big(\sum_{i=1}^{12} \phi(x \subseteq   \mathbb{S}^{\texttt{H}}_i) \ShiftLeft{} (12 + i)\big)
\label{scale_rep}
\end{equation}

\subsection{Feature Definitions}

Given a MIDI file $M$, for each note $n \in M$, $\texttt{ons}(n)$ returns the onset time of $n$ in ticks, $\texttt{dur}(n)$ returns the duration of $n$ in ticks, and $\texttt{pitch}(n)$ returns the pitch. An ordered set containing the unique onsets $O = \{\texttt{ons}(n) : n \in M\}$ is constructed, and the $i^{th}$ chord is the set of notes $\mathbb{C}^i = \{n : (\texttt{ons}(n) \leq O_i) \land (\texttt{ons}(n) + \texttt{dur}(n) > O_i)\}$. $\texttt{isOns}(\mathbb{C},n)$ and $\texttt{isTie}(\mathbb{C},n)$ are functions that return $1$ if $n$ is an onset or a tie respectively, and $0$ otherwise. The function $\texttt{pc}_i(\mathbb{C},n)$ returns $1$ if $n$ corresponds to the pitch class $i$ and $0$ otherwise. In order to simplify the feature definitions, we use Eq. \eqref{helper}, which accepts a chord $\mathbb{C}$ and a set of functions $F$, and only returns $1$ if there is an element in $X$ for which each $f \in F$ evaluates to 1. As a result, $\textbf{I}\big(\mathbb{C}, \{\texttt{isOns}, \texttt{pc}_i\}\big)$ is $1$ if there is a note $n \in \mathbb{C}$ that is an onset and is equivalent to the pitch class $i$.

\begin{subequations}
\medmuskip=0mu
\thinmuskip=0mu
\thickmuskip=1mu
\begin{flalign}
\texttt{pc}_i(\mathbb{C},n) = &
\begin{cases}
0, & \text{if}\ \texttt{pitch}(n) \mod \hspace{.25em} 12 \hspace{.25em} \equiv \hspace{.25em} i \\
1, & \text{otherwise}
\end{cases} \label{fa} \\
\texttt{isOns}(\mathbb{C},n) = &
\begin{cases}
0, & \text{if}\ \max(\{\texttt{ons}(n) : n \in \mathbb{C}\}) > \texttt{ons}(n) \\
1, & \text{otherwise}
\end{cases} \label{fb} \\
\texttt{isTie}(\mathbb{C},n) = & \hspace{.5em} 1 \hspace{.25em} - \hspace{.25em} \texttt{isOns}(\mathbb{C},n) \label{fc} \\
\textbf{I}(\mathbb{C}, F) = &
    \begin{cases}
      0, & \text{if}\ \max \big(\{\prod_{i=1}^{||F||} F_i(\mathbb{C},n) : n \in \mathbb{C}\}\big) < 1 \\
      1, & \text{otherwise}
    \end{cases}
    \label{helper}
\end{flalign}
\end{subequations}



Table \ref{feature_defs} provides formal definitions of all the features, where $\mathbb{C}^t$ denotes the $t^{th}$ chord, $\mathbb{M}^t$ denotes the $t^{th}$ melody pitch, $\mathbb{P}^t = \{\texttt{pitch}(n) : n \in \mathbb{C}^t\}$, $\mathbb{O}^t = \{\texttt{ons}(n) : n \in \mathbb{C}^t\}$, and $\mathbb{K}^t =  \{\texttt{pitch}(n) : (n \in \mathbb{C}^t) \land \texttt{isOns}(n)\}$. $\texttt{popcount}(\cdot)$ is a function that counts the number of set bits in an integer, $\texttt{pc}(x) = x \mod 12$ and $\texttt{pcc}(x) = |(x \mod 12) - 6|$. Dissonance is calculated using Stolzenburg's periodicity function \cite{stolzenburg2015harmony}, which we refer to as $\texttt{stol}(\cdot)$. Let $\texttt{diss}(\mathbb{P},\mathbb{T}) = \frac{1}{||\mathbb{T}||} \sum_{x \in \mathbb{T}} \texttt{stol}(\bar{\mathbb{P}}^x)$, where $\mathbb{P}$ and $\mathbb{T}$ are pitch sets, and $\bar{\mathbb{P}}^x = \{\mathbb{P}_i - x : \mathbb{P}_i \in \mathbb{P}\}$. $\texttt{voiceMotion}(\cdot)$ is a function that accepts two successive pitch sets $(\mathbb{P}^t, \mathbb{P}^{t+1})$ and returns an integer corresponding to the type of voice motion. $\texttt{tonnetzLength}(\cdot)$ is a function that accepts a pitch class set and returns the length of the shortest path through Tonnetz \cite{tonnetz} vertices containing each pitch class.

Each function is calculated for all valid values of $t$, resulting in a categorical distribution with unsigned 64-bit integers as the categories. For example, given a standard $4$-voice Bach chorale containing $m$ chords, the function \texttt{ChordSize} is calculated for $0 \leq t < m-2$, producing a categorical distribution with the categories $\{0,1,2,3,4\}$. In some cases, we weight values by chord duration, denoted by a $\star$ in the table. In the case that a function returns a set of values (\texttt{IntervalDist}), we combine the returned sets to form the categorical distribution. Since the number of categories $k$ grows exponentially large for some features (e.g., $\texttt{ChordShape}$), we restrict $k \leq 1000$ by ranking categories according to the number of samples they appear in, removing infrequently occurring categories.

\subsection{Implementation}

We implement the feature extraction tool in C++, using pybind11 \cite{pybind11} to create Python bindings. The Midifile library\footnote{\href{https://midifile.sapp.org/}{https://midifile.sapp.org/}} is used to parse MIDI files. 

\begin{table*}[t]
\centering
\resizebox{\textwidth}{!}{
\renewcommand{\arraystretch}{1.15}
\begin{tabular}{ c l | l | l}
& Feature Name & Function & Description \\
\hline
\parbox[t]{0.75mm}{\multirow{19}{*}{\rotatebox[origin=c]{90}{Chord}}} & ChordDissonance $\star$ & $\lfloor \texttt{diss}(\mathbb{K}^t, \mathbb{K}^t) \rfloor$ & the dissonance of onsets based on periodicity \cite{stolzenburg2015harmony} \\
\cline{2-4}
& ChordDistinctDurationRatio & $\big(1 \ShiftLeft{} ||\{\texttt{dur}(n) : n \in \mathbb{C}^t\}||\big)\,\,|\,\, 2^{||\mathbb{C}^t||}$ & the ratio of distinct note durations to chord size\\
\cline{2-4}
& ChordDuration & $\max(\mathbb{O}^{t+1}) - \max(\mathbb{O}^t)$ & the duration of a chord \\
\cline{2-4}
& ChordLowestInterval & $\min\big(\mathbb{P}^t \setminus \{\min(\mathbb{P}^t)\}\big) - \min(\mathbb{P}^t)$ & the difference between the lowest two notes \\
\cline{2-4}
& ChordOnset & $\big(\sum_{i=1}^{||\mathbb{C}^t||} (\texttt{isOns}(\mathbb{C}^t_i) \ShiftLeft{} (i-1)) \big)\,\,|\,\, 2^{||\mathbb{C}^t||}$ & an integer representing which notes are onsets \\
\cline{2-4}
& ChordOnsetPCD $\star$ & $\texttt{pcd}\big(\sum_{i=0}^{11} (\textbf{I}(\mathbb{C}^t,\{\texttt{isOns},\texttt{pc}_i\}) \ShiftLeft{} i)\big)$ & distinct pitch class set excluding ties\\
\cline{2-4}
& ChordOnsetRatio & $\big(1 \ShiftLeft{} \sum_{n \in \mathbb{C}^t} \texttt{isOns}(n)\big)\,\,|\,\, 2^{||\mathbb{C}^t||}$ & the ratio of onsets to chord size\\
\cline{2-4}
& ChordOnsetShape $\star$ & $\sum_{i=1}^{||\mathbb{C}^t||} (\texttt{isOns}(\mathbb{C}^t, \mathbb{C}^t_i)  \ShiftLeft{} (\mathbb{P}^t_i - \min(\mathbb{P}^t)))$ & piano roll type representation of onset pitches \\
\cline{2-4}
& ChordOnsetTiePCD $\star$ &  $\texttt{pcd}\big(\sum_{i=0}^{11} (\textbf{I}(\mathbb{C}^t,\{\texttt{isOns},\texttt{pc}_i\}) \ShiftLeft{} i)\big)\,+$ & concatenated distinct pitch class set of onsets  \\
& & $\texttt{pcd}\big(\sum_{i=0}^{11} (\textbf{I}(\mathbb{C}^t,\{\texttt{isTie},\texttt{pc}_i\}) \ShiftLeft{} i)\big) \ShiftLeft{} 12$ & and distinct pitch class set of ties \\
\cline{2-4}
& ChordOnsetTieReduced $\star$ & $\texttt{reduce}\big(\big(\sum_{i=0}^{11} (\textbf{I}(\mathbb{C}^t,\{\texttt{isOns},\texttt{pc}_i\}) \ShiftLeft{} i)\big)\,+ $ & concatenated pitch class set of onsets and pitch \\
& & $\big(\sum_{i=0}^{11} (\textbf{I}(\mathbb{C}^t,\{\texttt{isTie},\texttt{pc}_i\}\big) \ShiftLeft{} (12 + i))\big)\big)$ & class set of ties reduced using Eq. \eqref{rollb} \\
\cline{2-4}
& ChordPCD $\star$ & $\texttt{pcd}\big(\sum_{i=0}^{11} (\textbf{I}(\mathbb{C}^t,\{\texttt{pc}_i\}) \ShiftLeft{} i)\big)$ & distinct pitch class set\\
\cline{2-4}
& ChordPCDWBass $\star$ & $\texttt{pcd}\big(\sum_{i=0}^{11} (\textbf{I}(\mathbb{C}^t,\{\texttt{pc}_i\}) \ShiftLeft{} i)\big) + 2^{12 + \texttt{pc}(\min(\mathbb{P}^t))}$ & distinct pitch class set with bass pitch class \\
\cline{2-4}
& ChordPCSizeRatio & $\big(1 \ShiftLeft{} ||\{\texttt{pc}(p) : p \in \mathbb{P}^t\}||\big)\,\,|\,\,  2^{||\mathbb{P}^t||}$ & the ratio of distinct pitch classes to chord size\\
\cline{2-4}
& ChordRange $(\phi_1)$ & $\max(\mathbb{P}^t) - \min(\mathbb{P}^t)$ & the range of pitches in a chord \\
\cline{2-4}
& ChordShape $\star$ & $\sum_{p \in \mathbb{P}^t} (1 \ShiftLeft{} (p - \min(\mathbb{P}^t)))$ & piano roll type representation of chord pitches\\
\cline{2-4}
& ChordSize & $||\mathbb{C}^t||$ & the number of notes in a chord \\
\cline{2-4}
& ChordTonnetz $\star$ & $\texttt{tonnetzLength}(\{\texttt{pc}(x) : x \in \mathbb{P}^t\})$ & length of shortest path through Tonnetz \cite{tonnetz} vertices \\
\hline
\hline
\parbox[t]{0.75mm}{\multirow{12}{*}{\rotatebox[origin=c]{90}{Chord Transition}}} & ChordSizeNgram & $||\mathbb{C}^t|| + (||\mathbb{C}^{t+1}|| \ShiftLeft{} 8) + (||\mathbb{C}^{t+2}|| \ShiftLeft{} 16)$  & an $n$-gram of chord sizes $(n=3)$\\
\cline{2-4}
& ChordTranBassInterval & $\texttt{pc}(\min(\mathbb{P}^{t+1}) - \min(\mathbb{P}^t))$ & pitch class interval between two lowest notes\\
\cline{2-4}
& ChordTranDissonance & $\lfloor \texttt{diss}(\mathbb{P}^t, \mathbb{P}^{t+1}) \rfloor$ & the dissonance of intervals based on periodicity \cite{stolzenburg2015harmony} \\
\cline{2-4}
& ChordTranDistance & $|\min(\mathbb{P}^{t+1}) {-} \min(\mathbb{P}^t)| + |\max(\mathbb{P}^{t+1}) {-} \max(\mathbb{P}^t)|$ & approximated voice leading distance\\
\cline{2-4}
& ChordTranOuter & $\texttt{pc}(\phi_1(\mathbb{P}^t)) + (\texttt{pc}(\phi_1(\mathbb{P}^{t+1})) \ShiftLeft{} 8) +\,$ & pitch class transition using only the outer notes\\
& & $(\texttt{pc}(\min(\mathbb{P}^t) - \min(\mathbb{P}^{t+1})) \ShiftLeft{} 16)$ \\
\cline{2-4}
& ChordTranPCD & $\texttt{reduce}\big(\big(\sum_{i=0}^{11} (\textbf{I}(\mathbb{C}^t,\{\texttt{pc}_i\}) \ShiftLeft{} i)\big)\,+$ & transition between distinct pitch class sets \\
& & $\big(\sum_{i=0}^{11} (\textbf{I}(\mathbb{C}^{t+1},\{\texttt{pc}_i\}) \ShiftLeft{} (12 + i))\big), 24\big)$  \\
\cline{2-4}
& ChordTranRepeat & $(\prod_{n \in \mathbb{C}^t} \texttt{isOns}(n)) (\mathbb{P}^t = \mathbb{P}^{t+1})$ & chord repetition with onsets \\
\cline{2-4}
& ChordTranScaleDistance & $\texttt{popcount}\big(\texttt{sc}(\mathbb{P}^t) \BitXor{} \texttt{sc}(\mathbb{P}^{t+1})\big)$ & hamming distance between scale representations \\
\cline{2-4}
& ChordTranScaleUnion & $\texttt{popcount}\big(\texttt{sc}(\mathbb{P}^t)\,|\, \texttt{sc}(\mathbb{P}^{t+1})\big)$ & the union between scale representations\\
\cline{2-4}
& ChordTranVoiceMotion & $\texttt{voiceMotion}(\mathbb{P}^t, \mathbb{P}^{t+1})$ & type of voice motion (contrary, oblique, etc.) \\
\hline
\hline
\parbox[t]{0.75mm}{\multirow{2}{*}{\rotatebox[origin=c]{90}{Mel.}}} & MelodyNgram & $\sum_{i=0}^3 (\mathbb{M}_{t+i+1} - \mathbb{M}_{t+i} \mod 12) \ShiftLeft{} 8i$ & $n$-gram of melodic intervals $(n=3)$\\
\cline{2-4}
& MelodyPCD & $\texttt{pcd}\big(\sum_{i=0}^{11} \textbf{I}\big(\{\mathbb{M}^{t+i} : 0 \leq i < 5\},\{\texttt{pc}_i\}\big) \ShiftLeft{} i\big)$ & distinct pitch class of successive melody notes \\
\hline
\hline
\parbox[t]{0.75mm}{\multirow{2}{*}{\rotatebox[origin=c]{90}{Inter.}}} & IntervalClassDist & $\{\texttt{pcc}(p_i - p_j) : (p_j < p_i) \land (p_i, p_j \in \mathbb{P}^t \bigtimes \mathbb{P}^t)\}$ & interval class for each combination of chord pitches \\
\cline{2-4}
& IntervalDist & $\{\texttt{pc}(p_i - p_j) : (p_j < p_i) \land (p_i, p_j \in \mathbb{P}^t \bigtimes \mathbb{P}^t)\}$ & interval for each combination of chord pitches\\

\end{tabular}
\renewcommand{\arraystretch}{1}
}
\caption{Definitions for Chord features, Chord Transition features, Melody features (Mel.), and Interval features (Inter.). The $\star$ symbol indicates that a categorical distribution is weighted by chord duration.}
\label{feature_defs}
\end{table*}

\section{Similarity Computation}

In the most general sense, we are interested in measuring the similarity between a single MIDI file $\mathcal{X}$ and a corpus $\mathcal{C} = \{\mathcal{C}_1,...,\mathcal{C}_n\}$. We represent each MIDI file by applying a non-empty set of feature transformations $\mathcal{F} = \{f_1, ..., f_k\}$, producing a set of categorical distributions for each MIDI file. For each $f_i \in \mathcal{F}$, we aim to measure the similarity between a single categorical distribution $f_i(\mathcal{X})$ and a set of categorical distributions $f_i(\mathcal{C}) = \{f_i(\mathcal{C}_1),...,f_i(\mathcal{C}_n)\}$. Using a distance metric $\mathscr{D}$, the average similarity could be calculated $\frac{1}{n} \sum_{i=1}^{n} 1 - \mathscr{D}(f_i(\mathcal{C}_i), f_i(\mathcal{X}))$. However, this approach does not leverage information about the discriminative aspects of the entire corpus. The results in Experiment 1 demonstrate the deficiencies of this approach. Instead, we use Random Forests \cite{breiman2001random} to construct an embedding space before measuring the average similarity. Although neural networks are often ideal for learning embeddings, the time required to train $k$ neural networks is prohibitive for an online system.

Decision trees are commonly used to model complex data. When used to classify data, each terminal node represents a discrete class label, and an arbitrary input is classified based on the terminal node it reaches. Using a trained Random Forest, an input can be represented based on the terminal node it reaches in each decision tree. Given a Random Forest containing $N$ decision trees each with $L$ terminal nodes, an input can be represented as a vector $v \in \{0,1\}^{N \times L}$. To learn an embedding for a single feature transformation $f_i \in \mathcal{F}$, we train a Random Forest to discriminate between a collection of items $f_i(\mathcal{G}) = \{f_i(\mathcal{G}_1),..,f_i(\mathcal{G}_m)\}$ and a corpus $f_i(\mathcal{C}) = \{f_i(\mathcal{C}_1),...,f_i(\mathcal{C}_n)\}$. Concretely, each $f_i(\mathcal{G}_i) \in f_i(\mathcal{G})$ is given the label $0$, and each $f_i(\mathcal{C}_i) \in f_i(\mathcal{C})$ is given the label $1$. We refer to the vector produced for a sample $\mathcal{X}$ as $\textbf{R}^{\mathcal{G}, \mathcal{C}, f_i}_{\mathcal{X}}$. Breiman measures the similarity of two vectors using the dot product \cite{breiman2001random}. In order to weight each feature transformation $(f_i \in \mathcal{F})$ equally, we use cosine similarity (Eq. \eqref{cos}), which is simply the normalized dot product. The similarity between $\mathcal{X}$ and $\mathcal{C}$ with respect to a set of features $\mathcal{F}$ is computed using Eq. \eqref{dist}, which produces a scalar value on the range $[0,1]$.



\begin{subequations}
\begin{flalign}
    \texttt{cos}(X,Y) & = \dfrac{X \cdot Y}{\sqrt{\sum_{i=1}^N X_i^2} \sqrt{\sum_{i=1}^N Y_i^2}}
    \label{cos} \\
    S^{\mathcal{G}, \mathcal{C}, \mathcal{F}}_{\mathcal{X}} & = \dfrac{1}{||\mathcal{C}||||\mathcal{F}||} \sum_{c \in \mathcal{C}} \sum_{f \in \mathcal{F}} \texttt{cos}(\textbf{R}^{\mathcal{G}, \mathcal{C}, f}_{\mathcal{X}},\textbf{R}^{\mathcal{G}, \mathcal{C}, f}_c)
    \label{dist}
\end{flalign}
\end{subequations}



\section{Experiments}

In the following experiments, we train a Random Forest \cite{breiman2001random} using the scikit-learn python module \cite{scikit-learn}. We set the maximum tree depth at 5, the number of trees to 500, and measure the quality of the split using entropy. The class weight is balanced to be robust against size discrepancies between $\mathcal{C}$ and $\mathcal{G}$.

\subsection{Experiment 1 : Analytic Testing}


We test StyleRank with styles delineated by a single composer, and by an entire genre, using the Classical Archives MIDI dataset\footnote{\href{https://www.classicalarchives.com/midi.html}{https://www.classicalarchives.com/midi.html}}. In total there are 75 composers, and 6 musical genres. More details on the composition of the dataset can be found in the Appendix\footnote{\href{https://github.com/jeffreyjohnens/style\_rank/tree/master/appendix}{https://github.com/jeffreyjohnens/style\_rank/tree/master/appendix}}. We keep only one MIDI file per composition. Each MIDI file is represented as a list of pitches, sorted lexicographically according to onset and pitch. To compare two pieces, the Levenshtein distance \cite{levenshtein1966binary} is measured twice, once for the first 100 pitches in each piece, and once for the last 100 pitches. We eliminate pieces which have a Levenshtein distance less than $0.75$, after normalizing the distance on the range $[0,1]$. We choose this conservative value to ensure all duplicates are removed.

Given two styles $A = \{a_1,...,a_m\}$ and $B = \{b_1,...,b_n\}$, where $m=2n$, let $\mathcal{C} = \{a_i : 1 \leq i \leq n\}$, $\mathcal{G}_A = \{a_i : n < i \leq 2n\}$, $\mathcal{G}_B = B$, and $\mathcal{G} = \mathcal{G}_A \cup \mathcal{G}_B$. By construction $\mathcal{G} \cap \mathcal{C} = \varnothing$. We train a Random Forest and compare two distributions $x = [S^{\mathcal{G},\mathcal{C},\mathcal{F}}_g : g \in \mathcal{G}_A]$ and $y = [S^{\mathcal{G},\mathcal{C},\mathcal{F}}_g : g \in \mathcal{G}_B]$, where $\mathcal{F}$ denotes the set of features described in Table \ref{feature_defs}. Ideally, each value in $x$ should be larger than all values in $y$, since elements in $\mathcal{G}_A$ and $\mathcal{C}$ belong to the same style $(A)$. However, depending on the specificity of the style, there may be some degree of overlap between $A$ and $B$. In order to determine if there is a measurable difference between $x$ and $y$ we directly compare the means $(\bar{x} > \bar{y})$, and we calculate the $p$-value $(p^{\bar{x} > \bar{y}})$ for a One-Sided Mann-Whitney test \cite{mann1947} with the alternative hypothesis that $\bar{x} > \bar{y}$.

In cases where multiple statistical comparisons are performed, it is common practice to apply a correction to the raw $p$-values. The Bonferroni correction \cite{dunn1961} is calculated by dividing the desired level of significance $(\alpha = 0.05)$ by the number of comparisons. The Benjamini–Yekutieli procedure \cite{benjamini2001} controls the false discovery rate under arbitrary dependence assumptions, and is less conservative than the Bonferroni correction. Given $m$ null hypotheses and their corresponding $p$-values $P_1,...,P_m$, the $p$-values are sorted in ascending order. For a given level of significance, in our case $\alpha = 0.05$, reject the null hypothesis for the first $k$ values that satisfy $P_k \leq k\alpha / (m * c(m))$ where $c(m) = \sum_{i=1}^m 1/i$.

Table \ref{experiment_1_results} shows the results of 1000 trials, reporting the percentage of trials where $\bar{x} > \bar{y}$, and the percentage of trials where $p^{\bar{x} > \bar{y}}$ is significant, applying no correction $(\alpha = 0.05)$, the Benjamini–Yekutieli procedure (FDR), and the Bonferroni correction (Bon). We compare StyleRank against three distance measures, Cosine, Manhattan and Euclidean, replacing $S^{\mathcal{G},\mathcal{C},\mathcal{F}}_g$ with $\frac{1}{||\mathcal{C}||||\mathcal{F}||} \sum_{c\in\mathcal{C}} \sum_{f\in\mathcal{F}} 1 - \mathscr{D}(f(c), f(g))$.

\begin{table*}[t]
\centering
\resizebox{\textwidth}{!}{
\begin{tabular}{c@{\hspace{0.3cm}} c|*{4}{c@{\hspace{0.2cm}}}|*{4}{c@{\hspace{0.2cm}}}|*{4}{c@{\hspace{0.2cm}}}|*{4}{c@{\hspace{0.2cm}}}}
 & & \multicolumn{4}{c|}{StyleRank} & \multicolumn{4}{c|}{Cosine} & \multicolumn{4}{c|}{Manhattan} & \multicolumn{4}{c}{Euclidean}  \\
 & size & $\mu$ & Sig & FDR & Bon & $\mu$ & Sig & FDR & Bon & $\mu$ & Sig & FDR & Bon & $\mu$ & Sig & FDR & Bon \\ 
\hline
\parbox[t]{0.75mm}{\multirow{4}{*}{\rotatebox[origin=c]{90}{Composer}}} & 10 & 0.963 & 0.86 & 0.725 & 0.0 & 0.837 & 0.624 & 0.381 & 0.0 & 0.879 & 0.662 & 0.413 & 0.0 & 0.827 & 0.565 & 0.28 & 0.0 \\
& 25 & 0.951 & 0.888 & 0.807 & 0.609 & 0.808 & 0.583 & 0.422 & 0.24 & 0.793 & 0.578 & 0.415 & 0.244 & 0.729 & 0.532 & 0.363 & 0.226 \\
& 50 & 0.926 & 0.905 & 0.873 & 0.78 & 0.705 & 0.559 & 0.454 & 0.333 & 0.751 & 0.599 & 0.468 & 0.34 & 0.717 & 0.565 & 0.428 & 0.3 \\
& 100 & 1.0 & 0.986 & 0.973 & 0.951 & 0.713 & 0.636 & 0.59 & 0.515 & 0.723 & 0.633 & 0.568 & 0.486 & 0.715 & 0.626 & 0.571 & 0.504 \\
\hline
\parbox[t]{0.75mm}{\multirow{4}{*}{\rotatebox[origin=c]{90}{Genre}}} & 10 & 0.81 & 0.379 & 0.0 & 0.0 & 0.68 & 0.193 & 0.0 & 0.0 & 0.686 & 0.2 & 0.0 & 0.0 & 0.645 & 0.176 & 0.0 & 0.0 \\
& 25 & 0.867 & 0.578 & 0.376 & 0.198 & 0.729 & 0.348 & 0.084 & 0.038 & 0.74 & 0.374 & 0.053 & 0.021 & 0.691 & 0.298 & 0.06 & 0.022 \\
& 50 & 0.88 & 0.715 & 0.59 & 0.432 & 0.776 & 0.484 & 0.266 & 0.126 & 0.747 & 0.489 & 0.253 & 0.088 & 0.714 & 0.344 & 0.158 & 0.082 \\
& 100 & 0.927 & 0.847 & 0.774 & 0.671 & 0.766 & 0.555 & 0.406 & 0.265 & 0.755 & 0.566 & 0.44 & 0.284 & 0.785 & 0.462 & 0.269 & 0.178
\end{tabular}}
\caption{The normalized frequency over 1000 trials where $\bar{x} > \bar{y}$ $(\mu)$, $p^{\bar{x}>\bar{y}} < 0.05$ (Sig), $p^{\bar{x}>\bar{y}}$ is significant after applying the FDR correction (FDR), and $p^{\bar{x}>\bar{y}}$ is significant after applying the Bonferonni correction (Bon). Size denotes the size of the corpus $||\mathcal{C}|| = ||\mathcal{G}_A|| = ||\mathcal{G}_B||$.}
\label{experiment_1_results}
\end{table*}

\subsection{Experiment 2: Congruity with Human Perception}

In order to evaluate how well StyleRank correlates with human perception, we use data from the BachBot \cite{liang2017} experiment. In total, there were 5,967  participants, including 1329 novices, 2786 intermediate, 1341 advanced and 511 experts. Liang et al. generated 36 samples $(\mathcal{G})$ from a neural network trained on a collection of Bach Chorales $(\mathcal{C})$. Participants were asked to discriminate between a generated musical excerpt and an actual Bach chorale. They were each asked to complete 5 comparisons.

For each $g \in \mathcal{G}$, we count the number of times it was mistakenly classified as a Bach chorale $N^{\text{miss}}_g$, and the number of times it was correctly identified as computer generated $N^{\text{corr}}_g$. The raw count data can be found in the Appendix. We take the relative frequency of miss-classifications $T_g = N^{\text{miss}}_g / (N^{\text{miss}}_g + N^{\text{corr}}_g)$ as an indication of how similar $g$ is to the style of Bach's Chorales $(\mathcal{C})$. This results in ${36 \choose 2} = 630$ pairwise comparisons for which we have a ground truth ranking. Using a chi-square contingency test \cite{Pearson1992} we can measure the degree to which we are certain that there is a difference between two samples. We measure accuracy using Eq. \eqref{acc}, where $p_{ij}$ is the $p$-value for the chi-square contingency test comparing the counts for the $i^{th}$ and $j^{th}$ examples, $\phi(\cdot)$ is a function returning $1$ if the predicate $\cdot$ is true and $0$ otherwise, and $\alpha$ denotes the threshold for significance. 


\begin{subequations}
\begin{flalign}
    f(x,y) =
    \begin{cases}
        1, & \text{if}\ \phi\big(S^{\mathcal{G},\mathcal{C},\mathcal{F}}_x < S^{\mathcal{G},\mathcal{C},\mathcal{F}}_y\big) = \phi\big(T_x < T_y\big)\\
        0, & \text{otherwise}
    \end{cases}
    \label{ranking} \\
    \texttt{acc}(\mathcal{G},\mathcal{C},\alpha) = \dfrac{\sum_{i=1}^{||\mathcal{G}||} \sum_{j=i+1}^{||\mathcal{G}||} f(\mathcal{G}_i, \mathcal{G}_j) \phi(p_{ij} < \alpha)}{\sum_{i=1}^{||\mathcal{G}||} \sum_{j=i+1}^{||\mathcal{G}||} \phi(p_{ij} < \alpha)}
    \label{acc}
\end{flalign}
\end{subequations}

The results for Experiment 2 are presented in Table \ref{experiment_2_results}. We report the accuracy, calculated using Eq. \eqref{acc}, for a random ranking (Random), StyleRank with the jSymbolic \cite{mckay2018} features (jSymbolic), Log-likelihood (Loglik), and StyleRank. All the default features are extracted using jSymbolic, and features with zero standard deviation are removed. This results in a single feature vector with dimension of $453$, for which we train a single Random Forest. Using the Performance RNN \cite{magenta}, which was trained with the same representation and data as the original BachBot, we evaluate the negative log-likelihood $\mathscr{L}_g$ of each of the generated examples (loglik). To calculate the accuracy we simply replace the term $S^{\mathcal{G},\mathcal{C},\mathcal{F}}_X < S^{\mathcal{G},\mathcal{C},\mathcal{F}}_Y$ with $\mathscr{L}_X < \mathscr{L}_Y$ in Eq. \eqref{ranking}.

\begin{table*}[t]
\centering
\resizebox{\textwidth}{!}{
\begin{tabular}{c | *{4}{c@{\hspace{0.2cm}}} | *{4}{c@{\hspace{0.2cm}}}}
& \multicolumn{4}{c|}{Novice} & \multicolumn{4}{c}{Intermediate} \\
& $\alpha=\,$5.0 & $\alpha=\,$0.5 & $\alpha=\,$0.05 & $\alpha=\,$0.005 & $\alpha=\,$5.0 & $\alpha=\,$0.5 & $\alpha=\,$0.05 & $\alpha=\,$0.005 \\
\hline

Random & .482 $\pm$ .025 & .479 $\pm$ .031 & .466 $\pm$ .044 & .440 $\pm$ .062 & .500 $\pm$ .023 & .500 $\pm$ .026 & .502 $\pm$ .033 & .499 $\pm$ .037 \\
jSymbolic & .471 $\pm$ .006 & .463 $\pm$ .008 & .472 $\pm$ .012 & .491 $\pm$ .015 & .478 $\pm$ .011 & .474 $\pm$ .013 & .467 $\pm$ .014 & .456 $\pm$ .017 \\
Loglik & .629 $\pm$ .000 & .669 $\pm$ .000 & .764 $\pm$ .000 & .817 $\pm$ .000 & .654 $\pm$ .000 & .668 $\pm$ .000 & .690 $\pm$ .000 & .732 $\pm$ .000 \\
StyleRank & .716 $\pm$ .001 & .774 $\pm$ .002 & .855 $\pm$ .004 & .899 $\pm$ .005 & .702 $\pm$ .002 & .715 $\pm$ .002 & .758 $\pm$ .002 & .808 $\pm$ .002 \\

& \multicolumn{4}{c|}{Advanced} & \multicolumn{4}{c}{Expert} \\
\hline

Random & .511 $\pm$ .010 & .514 $\pm$ .013 & .512 $\pm$ .017 & .515 $\pm$ .019 & .493 $\pm$ .019 & .492 $\pm$ .025 & .492 $\pm$ .032 & .485 $\pm$ .038 \\
jSymbolic & .481 $\pm$ .011 & .480 $\pm$ .011 & .470 $\pm$ .014 & .474 $\pm$ .013 & .452 $\pm$ .008 & .449 $\pm$ .009 & .482 $\pm$ .012 & .464 $\pm$ .013 \\
Loglik & .673 $\pm$ .000 & .694 $\pm$ .000 & .730 $\pm$ .000 & .724 $\pm$ .000 & .657 $\pm$ .000 & .692 $\pm$ .000 & .741 $\pm$ .000 & .800 $\pm$ .000 \\
StyleRank & .718 $\pm$ .001 & .756 $\pm$ .001 & .806 $\pm$ .002 & .808 $\pm$ .002 & .692 $\pm$ .002 & .745 $\pm$ .003 & .821 $\pm$ .004 & .881 $\pm$ .005 \\

\end{tabular}}
\caption{The accuracy of each model, calculated using Eq. \eqref{acc}, with standard error calculated over 10 trials.}
\label{experiment_2_results}
\end{table*}

\section{Discussion}

Collectively, the results of both experiments demonstrate that StyleRank is robust to corpora of varying sizes, and highly correlated with human perception of stylistic similarity. In the Appendix, we expand Experiment 1 to demonstrate that StyleRank's performance is robust, even when the number of distinct styles in $\mathcal{G}$ is increased. In Experiment 1, there is a large difference between raw distance measures and StyleRank. This highlights the limitations of the approach described by Yang and Lerch, which uses euclidean distance to measure the distance between feature vectors \cite{Yang2018}. Although euclidean distance works well in low-dimensional settings, it does not scale well to high dimensions. In fact, it has been shown that Manhattan distance performs better than Euclidean distance in high dimensional settings \cite{Aggarwal}, which we also see in our own experimental results. Understandably, there is a decrease in performance when analyzing styles delineated by genre, as these styles have more variance, and are less consistent than the work of a single composer. Overall, these results demonstrate that StyleRank can proficiently rank MIDI files with different styles.

The results for Experiment 2 demonstrate that StyleRank is capable of making fine-grained distinctions between MIDI files that correspond with human perception of stylistic similarity. It is worth noting that participants found it difficult to discriminate between generated and human-composed samples in the BachBot experiment, evidenced by the average classification accuracy of novice (0.57), intermediate (0.64), advanced (0.68), and expert (0.71) participants \cite{liang2017}. Based on our experimental results, the jSymbolic \cite{mckay2018} feature set is no better at predicting rankings than a random model. This is likely due to the fact that high level features are not sufficiently discriminative for this task. In contrast to the jSymbolic feature set, our method involves full categorical distributions, which we believe are critical in measuring fine-grained differences. Importantly, there is a substantial difference between the accuracy of rankings based on log-likelihood and StyleRank. Interestingly, both log-likelihood and StyleRank best model high certainty $(\alpha = 0.005)$ comparisons made by self identified novices. This may be an artifact of increased variance as the number of ground truth comparisons decreases as $\alpha$ increases. 

It should be noted that participants in the BachBot experiment were not directly asked to rank samples according to their similarity to the style of Bach's chorales. We extrapolated a ranking from the number of times a sample was miss-classified, which is an indirect way of measuring stylistic similarity. However, since these rankings were based on a large sample size, we are confident that they are reflective of human perception.

\section{Application}

StyleRank can be used in a variety of settings. Importantly, we must note that there are no limitations on the composition of $\mathcal{G}$. For example, one could compare $k$ different sets with $\mathcal{G} = \{\mathcal{G}^1_i,...,\mathcal{G}^1_{n_1},\mathcal{G}^2_1,...,\mathcal{G}^2_{n_2},...,\mathcal{G}^k_1,...,\mathcal{G}^k_{n_k}\}$. First of all, the method can be use to rank samples generated by an SI system, based on their similarity to $\mathcal{C}$. StyleRank can be used to filter highly dissimilar samples automatically. Filtering is as simple as taking the samples $g \in \mathcal{G}$ with a similarity $S^{\mathcal{G},\mathcal{C},\mathcal{F}}_g$ above some threshold, and discarding the rest. Secondly, StyleRank can be used to rank models. Given $k$ models, let $\mathcal{G} = \{\mathcal{G}^1,...,\mathcal{G}^k\} = \{\mathcal{G}^1_i,...,\mathcal{G}^1_{n_1},...,\mathcal{G}^k_1,...,\mathcal{G}^k_{n_k}\}$, where $\mathcal{G}^i$ denotes the set of samples generated by the $i^{th}$ model. Then the distributions $x_i = [S^{\mathcal{G},\mathcal{C},\mathcal{F}}_g : g \in \mathcal{G}^i]$ can be compared using an appropriate statistical test. Third, the method can be used to isolate the specific features $f$ that deviate from the style delineated by $\mathcal{C}$ by comparing the distributions $x_f = [S^{\mathcal{G},\mathcal{C},f}_g : g \in \mathcal{G}]$ for each $f$ in a set of features $\mathcal{F}$. In addition, StyleRank can be used as a tool for musicologists to explore variations in style.

\section{Conclusion}

Quantifying musical stylistic similarity is a difficult task. We propose StyleRank, a method to rank individual MIDI files based on their similarity to an arbitrary style. Experimental evidence supports our approach, demonstrating that our method is robust, and is highly correlated with human perception of stylistic similarity. Future work involves applying this approach to other domains where SI systems are being developed. Additional features can be added to the current collection, in particular rhythm-based features, as the current collection is pitch-centric. Although we believe our experiments to be fairly comprehensive, continued validation of the proposed method on additional data is always beneficial.

\newpage

\section{Acknowledgments}

We acknowledge the support of the Natural Sciences and Engineering Research Council of Canada (NSERC), and the Helmut \& Hugo Eppich Family Graduate Scholarship.

\bibliography{ISMIRtemplate}


%
%
%
%

\end{document}